# Detection and Prevention of New and Unknown Malware using Honeypots

Shishir Kumar*, Durgesh Pant
*Jaypee Instuitute of Engineering & Technology
A-B Road, Raghogarh, Guna (MP) India-473226

**ABSTRACT** - Security has become ubiquitous in every domain today as newly emerging malware pose an ever-increasing perilous threat to systems. Consequently, honeypots are fast emerging as an indispensible forensic tool for the analysis of malicious network traffic. Honeypots can be considered to be traps for hackers and intruders and are generally deployed complimentary to Intrusion Detection Systems (IDS) and Intrusion Prevention Systems (IPS) in a network. They help system administrators perform a rigorous analysis of external and internal attacks on their networks. They are also used by security firms and research labs to capture the latest variants of malware. However, honeypots would serve a slightly different purpose in our proposed system. We intend to use honeypots for generating and broadcasting instant cures for new and unknown malware in a network. The cures which will be in the form of on-the-fly anti-malware signatures would spread in a fashion that is similar to the way malware spreads across networks. The most striking advantage of implementing this technology is that an effective initial control can be exercised on malware. Proposed system would be capable of providing cures for new fatal viruses which have not yet been discovered by prime security firms of the world.

*Keywords - Honeypots, Malware, Security, Anti-malware signatures, Malware detection*

## 1. INTRODUCTION

Several security firms across the world are busy preparing patches and cures for the plethora of malware existent today. But the fact remains that for every cure created for a malware, a subtle variant of the same malware is created that bypasses all the latest security patches thereby nullifying all the hard work and effort put in to counter them. To make things worse, malware is becoming smarter everyday and polymorphic malware are the latest entrants in this calamitous game of defeating the opponent. These malware are capable of self-reproduction, each instance of which adopts a completely different identity from its parent. In several recent malware outbreaks, especially pertaining to fatal viruses, considerable damage had already been done to thousands of systems and networks worldwide before the viruses were discovered and patches were released to invalidate them. Thus, we must create an anti-malware mechanism that is capable of creating and spreading cure for new variants of malware at the same speed and in the same fashion that malware adopts for spreading across networks.

Our research proposal aims at providing a solution to the above described problem. We intend to use honeypots as a tool to capture new and unknown malware. Once detected, our honeypot will create on-the-fly anti-malware signatures and broadcast them throughout the network being guarded by it. Individual hosts will then update their anti-malware signatures and thus remain protected against any threat posed to them by fatal malware. The entire process of detection of new malware and the creation and broadcast of a cure for it on a particular network would ideally be a matter of a few seconds or minutes. This is obviously much quicker than waiting for major security firms to first discover the new malware and then release patches for them. The following paragraphs describe the approach that we would adopt in order to realize this technology.

## 2. RELATED WORK

Detection of new variants of malware is among the major research activities that are carried out by security firms in the IT industry. A majority of anti-virus firms and research organizations use honeypots to capture variants of existing as well as new malware. They work upon the acquired data and binary samples to produce patches for defense against malware threats. For instance, Avira, an anti-virus firm uses distributed honeynets to gather binary samples of new malware from networks and subnets spread all over the globe [9]. The combination of such data from a variety of sources proves to be extremely useful in creating efficient security patches.

Another attempt to use honeypots with anti-virus capabilities was made by IBM in 2005. The project was named Anti-Virus in the Wild [8] and a presentation on the same was made in the Virus Bulletin Conference in October 2005 at Dublin. It consisted of a network of thirteen virtual Windows XP systems (each serving as a honeypot) realized using four real machines. Each honeypot had one popular anti-virus installed on it with no other network protection enabled. A single firewall guarded the entire network as a whole. The main aim of the project was to compare the efficiency of the thirteen most popular anti-virus software in the industry.

The projects mentioned above have used honeypots in accordance with malware detection but their main aim was to detect and gather binary samples of variants of existing and new malware, especially viruses. Our research initiative takes this mechanism one step ahead as we intend to generate automated cures for malware that will be detected by our honeypot. These cures would then spread across the network guarded by our honeypot adopting the same mechanism that malware uses to propagate itself, thereby nullifying the satanic intentions of malware creators.

A concept similar to the one proposed by us has been presented by Frank Castaneda, Emre Can Sezery, and Jun Xuy in their paper 'Worm Vs. Worm' [1]. They have proposed a method that transforms a malicious worm into an anti-worm to disinfect its original and then evaluated this method using the CodeRed, Blaster and Slammer worms. In addition, researchers at Columbia University, New York have are also working on a project named 'A Network Worm Vaccine Architecture' which presents an architecture that is very much in conjunction to that proposed by us [7].

## 3. PROPOSED APPROACH

The most important consideration in view of our solution is that it will be completely network based. Our proposed anti-malware system will be divided into two major parts: A honeypot server and a thin client. An instance of our thin client will be installed on each host present in the network being guarded by the honeypot server.

The honeypot server can be considered to be the protagonist of our anti-malware system. This will be a high-interaction research honeypot and shall be responsible for detecting all kinds of malware on the network under surveillance. To prepare our honeypot for this task, we will have to equip it with robust malware-detection mechanism. In the first version of our system, we would restrict ourselves only to pattern-based detection of malware. Heuristics and new technologies like buffer overflow protection and network port blocking would be considered for inclusion into the malware detection mechanism after the first version of our system becomes efficient and robust.

Apart from the ability to detect malware, our honeypot would in the future also be capable of generating cures in the form of anti-malware packets for every malware detected by it. An anti-malware packet will consist of an anti-malware signature, a set of operating system undo operations and a harmless anti-worm created out of the original malware. The signatures would be generated on-the-fly as soon as a new malware is detected and then inserted into an anti-malware packet. Subsequently, these packets would be broadcasted to all the hosts present on the network being guarded by our honeypot.

Several technologies would be involved in the creation of a mechanism for the detection of malware and subsequently for generating anti-malware signatures for them. Phenomenal research work is currently underway in these areas. We will tweak these existing technologies before incorporation into our honeypot. This would ensure that our anti-malware system produces efficient output to the maximum possible extent with a minimum number of false positives.

### 3.1 THE MALWARE DETECTION PHASE

The first phase of our system involves monitoring of all network traffic that passes through our honeypot and detection of code that possibly may be malicious in nature. Suspected pieces of malicious code will then be redirected to a virtual machine installed on the honeypot itself. Program monitoring tools in the virtual machine would then perform analysis on the code to ensure if it really is malicious in nature or not. If the code is indeed of malicious nature, then the virtual machine would further monitor its activities and create a log of the changes that the code makes to the operating system. Therefore, in order to reduce overheads on the honeypot as well as the virtual machine, it is important for the malware detection algorithms that will be used in this phase to be extremely robust. These algorithms should also produce as few false positives and false negatives as possible, although it is not practically possible to achieve hundred percent accuracy.

Most of the existing malware detection algorithms rely on pre-defined malware signature databases to scan network traffic for traces of patterns which are known to them. However, the aim of our system is to protect a network against new and unknown malware. The use of pre-defined signature based detection algorithms would not produce satisfactory results in our case. Therefore, we need to device a new algorithm that would enable us to detect new and unknown patterns that may potentially be malicious in nature. The following paragraphs define and explain the approach adopted by us to accomplish this task.

### 3.2 DETECTION OF MALICIOUS TRAFFIC

In order to detect malicious content in network traffic, we must use background services deployed on the honeypot as well as the capture device. Windows services are the best examples of background services. The capture device should be used exclusively to capture live network traffic in its native format and the other machine should be the honeypot where the actual detection process takes place.

We propose a malware detector service which analyses the captured network traffic, extracts relevant network traffic patterns from it and then uses several statistical methods to determine which patterns may be malicious in nature. It would help if the captured packets are converted into packet sets before analysis. This improves the statistical analysis done on the relevant patterns.

The Bloom filter, conceived by Burton H. Bloom in 1970, is a space-efficient probabilistic data structure that is used to test whether an element is a member of a set. False positives are possible, but false negatives are not. Elements can be added to the set, but not removed (though this can be addressed with a counting filter). The more elements that are added to the set, the larger the probability of false positives.

An empty Bloom filter is a bit array of m bits, all set to 0. There must also be k different hash functions defined, each of which maps a key value to one of the m array positions.

The probability of getting a collision in the Bloom Filter varies with the following equation:

$$\left(1 - \left(1 - \frac{1}{m}\right)^{kn}\right)^k \approx \left(1 - e^{-kn/m}\right)^k$$

We propose to use a Bloom Filter with 10,000 bits to find intersecting patterns between two packets. Assuming that each packet size would be approximately 1500 bytes, the probability of encountering a collision would be as low as $6.1 \times 10^{-4}$. With such a low collision probability, the hashing algorithm described below can be deemed robust enough for this application.

We would be using a simple hashing algorithm to generate hash codes for patterns that are relevant to the malware detection process. The main advantage of using hashing to compare patterns over direct string comparison lies in the speed of comparison. Let us consider a pattern ($C_1 C_2 \ldots C_n$) to be hashed. Let the lower limit of pattern length be K and the number of bits in our Bloom Filter be M. Then,

$$H = c_1 q^{k-1} + c_2 q^{k-2} + \ldots + c_{k-1} q + c_k \; (\text{Mod } M)$$

Each packet set contains several packets captured in a sequence. This ensures that every byte that is transferred on a network is also stored in the captured packets. Our next task is to extract from these packet sets, all the patterns that are relevant to our detection process. Relevance may be judged on the basis of several parameters, the most common one being length. Extremely short patterns (those with length less than 10 bytes) have the tendency of occurring with high frequencies in malware as well as legitimate traffic rendering such patterns useless for our algorithm. In contrast, studies show that the length of a typical malware code does not exceed a certain amount to ensure its fast propagation over the Internet. This implies that we need to look for patterns with length that is optimal for malware identification by setting lower and upper thresholds.

Fang Hao, Murali Kodialam, T.V. Lakshman and Hui Zhang state in their publication 'Fast Payload-Based Flow Estimation for Traffic Monitoring and Network Security' [3] that using a lower threshold in the range of 20 bytes to 40 bytes can be considered to be a good tradeoff to start with. On the other hand, setting the upper threshold is not a simple task and is another research problem as we can never be sure of the maximum code length that malware developers may decide upon. Setting a rigid upper threshold may also allow smaller suspicious patterns to pass through our detection mechanism.

Once we decide a minimum pattern length to be searched for, we start scanning each captured packet for patterns that might be of interest to us. We accomplish this task in groups of two packets each in a way similar to that described in [3]. Note that each packet set analyzed by us is a separate file which contains an array of packets within itself. Thus at any point of time, our algorithm analyses a packet set by analyzing its constitutional packets in groups of two to find patterns that are common to both the packets. This procedure can be considered to be a modification of the well-known Longest Common Substring (LCS) problem. We find patterns of all lengths that are common to both the packets currently being analyzed but with the restriction that the patterns must be of a minimum length as specified in our algorithm.

Let us consider two packets $P_1$ and $P_2$ of unequal length with the following contents:

Packet $P_1$

| A | B | C | D | E | F | G | H | I | J | K |
|---|---|---|---|---|---|---|---|---|---|---|

Packet $P_2$

| A | M | N | B | C | D | O | P | Q | G | H | I | J | R |
|---|---|---|---|---|---|---|---|---|---|---|---|---|---|

The common patterns in the above two packets have been shaded appropriately. Assuming that we have to find all the common patterns in $P_1$ and $P_2$ with a minimum length of 3 bytes (denoted as 3-patterns), the following two patterns would be of interest to us. We would discard the pattern 'a' because it does not satisfy our minimum length criteria.

| G | H | I | J |
|---|---|---|---|

| B | C | D |
|---|---|---|

The above described procedure for finding patterns relevant to our algorithm is repeated several times for all the packet sets in a corpus captured by the Packet Capture Service. Every time a new pattern is discovered, we store it in a table called the Coincidence Count Table along with its hash code and the number of times it has been discovered in the captured packet corpus. If a particular pattern is discovered more than once, its count in the Coincidence Count Table is incremented accordingly. The coincidence count of each pattern is also used to get an estimate of the fraction of captured packets in that packet set in which that particular pattern occurs.

Let $f_Q$ denote the fraction of the packets containing a pattern Q (assuming that we have N packets per packet set) and let S(Q) be the coincidence count of that pattern. Then,

$$f_Q = \sqrt{\frac{S(Q)}{N}}$$

We create one coincidence count table for each packet set. Therefore, if we have ten packet sets per corpus, we will have ten coincidence count tables associated with each corpus.

After all the packet sets are searched for patterns of interest to us, our coincidence count tables get ready for statistical analysis. We lay more emphasis on those patterns that have a higher value of $f_Q$. Intuitively, we can set a threshold for $f_Q$ above which all the patterns need to be analyzed in detail to see the changes they make to an OS. However, proceeding in this fashion may lead to a high rate of false positives, something that we wish to reduce in order to improve the overall performance of our system. We therefore use a statistical technique called Inverse Distribution followed by Standard Deviation to further analyze the $f_Q$ value of relevant patterns.

An efficient algorithm for performing inverse distribution analysis has been presented by Vijay Karamcheti, Davi Geiger, Zvi Kedem and S. Muthukrishnan in [2]. We wish to use a similar technique to reduce the number of false positives generated by our detection scheme.

## 4. EXPERIMENTAL RESULTS

To test our detection algorithm, we created three different windows services: The Packet Capture Service, The Corpus Receiver Service and The Malware Detector Service. The former service runs on the capture device while the other two run on the honeypot. We created a trivial FTP for transferring captured packets from the capture device to the honeypot. As we stated earlier, the captured packets are converted into packet sets. Several packet sets are clubbed together to form a corpus which is sent across to the honeypot using the trivial FTP. For testing purposes, we have decided to enclose 100 packets in one packet set and 10 packet sets in one corpus. The capture device we have chosen is a server which has global access to the internet. This would enable us to experiment with real network data for better results.

The Malware Detector Service on the honeypot periodically applies the algorithms described above in order to detect possibly malicious traffic in the captured packet sets. We have set the periodicity of this service as one hour for testing purposes. All suspected strings are appended to a file which is stored in the honeypot.

We have operated this setup for about four weeks and manually analyzed the type of network traffic that is deemed malicious by it. Many of the detected strings consist of java script and plain HTML code. While it is unlikely for plain HTML code to be malicious in nature, java script can cause a lot of damage to systems and networks if intended to do so. We also spread the test virus EICAR on the network and it was detected by our system. However, further analysis is required in order to correctly identify malicious items. Our suggestions of future work in the next section can provide some valuable tips for continuing research on this topic.

## 5. FUTURE WORK

Our anti-malware system aims at providing complete protection to networks against new and unknown viruses. However, it is a well-known fact that malware is ever-evolving and malware authors learn new tricks everyday to help them easily evade the existing security mechanisms. Therefore, the task of first detecting unknown suspicious network activity, then analyzing them to see the changes they make to our operating systems and finally generating cures for them are in themselves individual research topics. To the best of our knowledge, no such anti-malware system exists in the world as of today. Hence, this proposal bears tremendous scope for future work in the form of enhancements and optimizations to the work done by us.

We have limited this paper to the detection of possibly malicious network traffic. However, the detected mal-strings must be further analyzed to ensure that they really are malware and if that is true, then what changes they make to the operating system. This is a vast research topic which requires in-depth knowledge of binary instruction sets and also how an operating system functions. Although, we have already begun working on this front, a lot more needs to be accomplished before our analysis architecture can be made robust enough. As a small step towards achieving this goal, Amit Vasudevan and Ramesh Yerraballi have proposed powerful dynamic fine-grained malicious code analysis frameworks like Cobra [4] and SPiKE [6], to combat malware that are becoming increasingly hard to analyze. DOME [5] is another host-based technique for detecting several general classes of malicious code in software executables. Enhanced versions of such frameworks can be extremely helpful in analysis of malware.

In addition, we can also install malware aware thin clients capable of scanning systems for malware known to them on every host in our network. Each instance of our thin client will maintain a database of anti-malware signatures. As soon as a new signature is broadcasted over the network by the honeypot server, the thin clients would update their malware database with the incoming packet. This would be followed by a malware scan of each node by thin clients. If the signature of a malware present on our network exists in the database of the thin clients, it would subsequently be either deleted, healed or quarantined by the thin clients, as demanded by the situation. Moreover, since the database of the thin clients would consist of a very small number of definitions as compared to commercially available anti-virus and anti-spyware software, the total amount of time taken to scan a system would be considerably less as compared to other scanners. This will result in preservation of valuable computing resources.

However, despite the several advantages of our anti-malware system, it contains certain pitfalls which we would overcome in future versions of our system. The thin clients installed on each node of a network should be able to initialize a scan irrespective of whether the honeypot server detects a malware and broadcasts its anti-malware definition over the network or not. Thus even if a known malware is injected into the network via secondary media, its detection and removal can be done with the help of thin clients. This will enable thin clients to provide real time malware protection and external triggers will not be required to initiate scans. Additionally, all thin clients should be capable of triggering network scans whenever they detect a malware. This will expedite the process of malware detection and removal.

Another major drawback of our system as mentioned earlier is that currently our system relies entirely on pattern-based detection of malware. After the current version of our system becomes efficient and robust, we would also include better malware detection technologies like heuristics, buffer overflow protection and network port blocking into the malware detection mechanism of our honeypot. This will further help reduce the number of false positives generated during the detection phase by our honeypot.

## 6. CONCLUSION

Our anti-malware system has several advantages which include protection against new and unknown malware threats which have not yet been discovered by major security firms. This would also hold true for less popular yet harmful malware which are not globalized but confined to specific geographical locations. Moreover, since the size of the anti-malware signature database would be very small as compared to those contained by popular anti-virus and anti-spyware products, system scans performed by our thin clients would take very little time consequently saving valuable system resources. In addition, our honeypot would also serve as a valuable research tool for the analysis of new and upcoming malware production and distribution techniques.

With the help of our proposed system, we have introduced the idea of using honeypots for the detection and prevention of new and unknown malware, especially localized yet harmful malware and also those for which security patches have not yet been released by major security vendors. We have suggested how our proposal is an extension of the work that honeypots are being used currently for. In order words, while honeypots are basically used for gathering binary samples of newly evolving malware, our anti-malware system would use honeypots for detecting new and unknown malware. Furthermore, our honeypot can also help in generating anti-malware signatures for the possibly malicious code detected by it. These signatures would be broadcasted over the network under surveillance and subsequently would be used by thin clients to clean malware from individual hosts. The algorithm which we propose to be used by our honeypot for detecting possible malware has also been illustrated in detail. Finally we listed the advantages and disadvantages of our system.